\documentclass[journal]{IEEEtran}

\usepackage{amsmath}
\usepackage{amssymb}
\usepackage{booktabs}
\usepackage{multirow}
\usepackage{algorithm}  
\usepackage{algpseudocode}  

\usepackage{pifont}
\usepackage{graphicx}
\graphicspath{{figures}}
\usepackage{bbm}
\usepackage{float}
\usepackage[caption=false,font=footnotesize,labelfont=rm,textfont=rm]{subfig}
\usepackage{cite}
\usepackage{enumerate}

\newtheorem{theorem}{Theorem}
\newtheorem{lemma}{Lemma}
\newtheorem{corollary}{Corollary}
\newtheorem{remark}{Remark}

\usepackage{color}
\usepackage{threeparttable}

\begin{document}
\title{A Novel Bilateral Energy Trading Mechanism for Electricity Markets with Numerous Prosumers}
\author{Bing Liu,
        Furan Xie
        and Li Chai*
\thanks{This work is supported by National Natural Science Foundation of China under grant number 62173259.}
\thanks{Bing Liu and Furan Xie are with the Engineering Research Center of Metallurgical Automation and Measurement Technology, Wuhan University of Science and Technology, Wuhan 430081, China (e-mail: liubing17@wust.edu.cn; xiefuran328@wust.edu.cn).}
\thanks{Li Chai is with the College of Control Science and Engineering, Zhejiang University, Hangzhou 310027, China (e-mail: chaili@zju.edu.cn).}
\thanks{*Corresponding Author.}}
\maketitle
\begin{abstract}
With the rapid development of distributed energy resources, increasing number of residential and commercial users have been switched from pure electricity consumers to prosumers that can both consume and produce energy. To properly manage these emerging  prosumers, a peer-to-peer (P2P) electricity market has been explored and extensively studied. In this paper, a scalable energy management mechanism is proposed for the P2P electricity market. First, the multi-bilateral economic dispatch problem that maximizes social welfare is formulated, taking into account product differentiation and network constraints. Then, an energy management mechanism is devised to improve the scalability from two aspects: (i) an accelerated distributed clearing algorithm with less exchanged information and faster convergence rate. (ii) a novel selection strategy to reduce the amount of computation and communication per player. Finally, the convergence rate of the proposed accelerated algorithm is given, and the proposed selection strategy is illustrated through a Monte Carlo simulation experiment.

\end{abstract}


%
\IEEEpeerreviewmaketitle

\section{Introduction}
%
%
%
%
\IEEEPARstart{I}{n} the past, electricity generation has always relied on fossil fuels, contributing about one-third of annual carbon emissions \cite{2}. With aggravation of global warming, more and more countries have set up ``carbon peak'' and ``carbon neutral'' action goals to gradually reduce carbon emissions. To achieve the stated action goals, new energy sources, mainly solar energy and wind energy, have received unprecedented attention and rapid development. So in recent years, more and more residential and commercial users have been equipped with photovoltaic panels, wind turbines, and distributed energy resources, becoming the so-called prosumers that both consume and produce electricity. These emerging prosumers are geographically decentralized and more willing to actively participate in the electricity market for additional revenue \cite{4}.

To properly manage these prosumers, a so-called peer-to-peer (P2P) electricity market has been proposed and received widespread attentions \cite{48}. In such an electricity market, each player can directly trade energy with other players at mutually satisfactory prices. There are three main energy trading strategies: game theory, double auction, and optimization-based theory \cite{10}. Game theory analyzes the decision-making process of market players in which each player's action depends on and influences other players' actions \cite{51}. In the double auction, buyers and sellers submit offers and bids to the auctioneer, and the auctioneer facilitates market transactions by coordinating the prices of both step by step \cite{47}. However, neither game theory nor double auction can guarantee optimal energy allocation. In contrast, due to optimality, robustness, scalability, and security concerns, an increasing number of research has focused on distributed optimization-based methods to solve the energy trading problem \cite{22, 50, 49}.

In the above work, electricity is regarded as a homogenized good and priced under uniform pricing in the forward market. However, in practice, consumers should be allowed to express their preferences on the type and quality of electricity, which may result in product differentiation and further price differentiation \cite{A-1}. 
In \cite{A-1}, a multi-bilateral energy trading framework that allows product differentiation is proposed for the P2P electricity market. The optimal solution of the multi-bilateral energy trading problem with product differentiation is analyzed in \cite{42}. Product differentiation is further enriched and developed in \cite{31}, in which the energy is classified to meet individual preferences based on attributes of its source, such as generation technology, location in the network and the owner's reputation. In addition, product differentiation can also be used to implement consumer-related costs, such as taxes and network usage fees \cite{33,38}.

On the other hand, in the P2P electricity market, since each player trades directly with other players, the number of transactions and optimization variables increases exponentially with the number of market players. This will vastly increase the computation and communication burdens for each player, presenting serious challenges to the scalability of the market. At present, the scalability can be improved by either reducing the exchanged information between players or the number of trading partners per player. The former can reduce the amount of communication while the latter can reduce both the amount of communication and computation. For example, in the P2P energy trading scheme proposed in \cite{A-2}, only price or energy information is transmitted between producers and consumers, which reduces the exchanged information by half compared to the RCI approach in \cite{A-1}. In \cite{28}, the concept of node coloring is presented to improve scalability, in which each player only trades with players of the same color. In \cite{13}, an additional adaptive segmentation procedure is proposed in which all players are grouped into different segments such that each player only trades with players within the same segment. It also enhances the scalability by reducing the number of trading partners per player. However, this approach of reducing trading partners may cause the loss of social welfare due to artificially limiting possible transactions. In addition, the adaptive segmentation method in \cite{13} requires to additionally slove a strongly NP-hard problem. In this paper, besides reducing the exchanged information, the scalability is enhanced by directly improving the convergence rate of the clearing algorithm. In addition, a simple selection strategy is presented to choose efficient trading partners for each consumer, which can further improve the scalability while keep the optimal social welfare as much as possible. In summary, the main contributions of this paper are as follows:
\begin{itemize}
\item [1)]
A distributed clearing algorithm is proposed to solve the multi-bilateral economic dispatch problem with product differentiation. Compared with other methods, the proposed algorithm has less exchanged information and faster convergence rate.
\item [2)]
A novel selection strategy based on consumer preferences is presented, which can reduce the number of inefficient trading partners for consumers and thus reduce the computational and communication burdens per player.

\item [3)]
A detailed theoretical analysis of the convergence rate of the proposed clearing algorithm is given, and the effectiveness of the proposed selection strategy is illustrated by a Monte Carlo simulation experiment.

\end{itemize}

The rest of this paper is structured as follows. Section \uppercase\expandafter{\romannumeral2} formulates a multi-bilateral economic dispatch problem for maximizing the social welfare. Section \uppercase\expandafter{\romannumeral3} proposes a scalable energy trading mechanism. Section \uppercase\expandafter{\romannumeral4} gives the main theoretical results. Section \uppercase\expandafter{\romannumeral5} conducts numerical simulations. Section \uppercase\expandafter{\romannumeral6} gathers conclusions of the paper.

\section{Problem Formulation}
In the P2P electricity market, prosumers can trade directly with other prosumers to sell or buy energy. The role of a prosumer, producer or consumer, depends on its net power within a single time slot. Our focus is on the problem of maximizing the welfare of all producers and consumers.

Consider a P2P electricity market with $n$ prosumers. The communication network among prosumers is represented by an undirected graph $\mathcal{G}=(\mathcal{N},\mathcal{E})$, where $\mathcal{N}$ is the set of prosumers and $\mathcal{E} \subseteq \mathcal{N}\times \mathcal{N}$ is the set of communication edges. The producers set and the consumers set are defined as $\mathcal{N}_P=\{1,2, ..., n_p\}$ and $\mathcal{N}_C=\{1,2, ..., n_c\}$, satisfying $\mathcal{N}_P\bigcup\mathcal{N}_C=\mathcal{N}$, $\mathcal{N}_P\bigcap\mathcal{N}_C=\emptyset$ and $n_p+n_c=n$. 
Let $\mathcal{N}_i$ and $\mathcal{N}_j$ be the sets of neighbors of producer $i$ and consumer $j$ with capacity $N_i$ and $N_j$, respectively. A radial distribution network for transmitting energy is composed of a set of buses $\mathcal{N}_B = \{0, 1, \dots, n\}$ and a set of distribution lines $\mathcal{N}_L$. The root of the radial network (bus 0) is considered as the slack bus. Except for bus 0, each bus corresponds to one prosumer. In addition, all vectors are regarded as column vectors.

\subsection{Welfare Function of Producers}
The generation cost of producer $i$ is associated with its total generation. Define the sold energy of producer $i$ to consumer $j$ as $x_{ij}$, then the total generation of producer $i$ is $x_i=\sum_{j\in\mathcal{N}_i}x_{ij}$.
Then, the cost function of producer $i$ can be represented as \cite{A-1,49}
\begin{gather}
\begin{aligned}
\label{fun_c}
&C_i(x_i)=\frac{1}{2}a_ix_i^2+b_ix_i+c_i,\\
\text{with}&\quad x_i^{min}\le x_i\le x_i^{max}, \quad\forall i\in\mathcal{N}_P,
\end{aligned}
\end{gather}
where $a_i$, $b_i$, and $c_i$ are positive cost coefficients of producer $i$. $x_i^{min}$ and $x_i^{max}$ are the minimum and maximum generation of producer $i$, respectively. Note that for new energy producers, since its generation is a fixed value in a particular time slot, the constraint in (\ref{fun_c}) can be remodelled by setting $x_i^{min}$ equal to $x_i^{max}$. Generally speaking, it is reasonable to model classical producers' cost function as a quadratic form, while that of new energy producers is currently unclear. However, since its generation is a fixed value, the generation cost of new energy producers is a constant in a single time slot, which will not affect the optimal energy allocation. For the sake of brevity, the cost function of producers are uniformly modeled as a quadratic form.

The welfare function of producer $i$ is composed of the generation cost and the trading profit. Suppose the trading price between producer $i$ and consumer $j$ is $\lambda_{ij}$. Then, the welfare function of producer $i$ is obtained
\begin{equation}
\begin{aligned}
\label{eq.wp}
WP_i &= \sum_{j\in\mathcal{N}_i}\lambda_{ij}x_{ij}-C_i(x_i)=\lambda_i^T\mathbf{x}_i - C_i(x_i),\\
&\text{with}\quad x_i^{min}\le x_i\le x_i^{max}, \quad\forall i\in\mathcal{N}_P,
\end{aligned}
\end{equation}
where $\lambda_{i}=(\lambda_{ij})_{j\in\mathcal{N}_i}$ and $\mathbf{x}_{i}=(x_{ij})_{j\in\mathcal{N}_i}$ are column vectors containing the trading price and energy of producer $i$ with its neighbors, respectively.


\subsection{Welfare Function of Consumers}
For each consumer $j\in\mathcal{N}_C$, consuming a certain amount of energy will get corresponding convenience and satisfaction. This kind of welfare obtained through consumption can be expressed by the utility function in microeconomics \cite{34}. Define the energy purchased by consumer $j$ from producer $i$ as $y_{ji}$, and the total energy purchased by consumer $j$ is $y_j=\sum_{i\in\mathcal{N}_j}y_{ji}$. Then, the utility function with linearly decreasing marginal benefit can be formulated as \cite{15}
\begin{equation}
\begin{aligned}
\label{fun_u}
U_j&(y_j) =
\begin{cases}
\omega_jy_j-\frac{\delta_j}{2}y_j^2 &,\quad 0\le y_j\le\frac{\omega_j}{\delta_j},
\\\frac{\omega_j^2}{2\delta_j} &,\quad y_j\ge \frac{\omega_j}{\delta_j},
\end{cases}\\
&\text{with}\quad y_j^{min}\le y_j\le y_j^{max}, \quad\forall j\in\mathcal{N}_C,
\end{aligned}
\end{equation}
where $\omega_j$ represents the value of electricity for consumer $j$. $\delta_j$ is a predefined parameter. $y_j^{min}$ and $y_j^{max}$ are the minimum and maximum demand of consumer $j$, respectively. In particular, the utility function (\ref{fun_u}) is modeled for consumers with flexible loads, such as smart appliances and plug-in electric vehicles, etc. For consumers with inflexible loads, the utility function $(\ref{fun_u})$ is still applicable by setting $y_j^{min}$ and $y_j^{max}$ equal to each other.  In addition to the above quadratic form, any function with the property of diminishing marginal benefit can be used to quantify the welfare that consumers receive from consumption.


In the P2P market, the welfare of consumer $j$ is not only composed of utility function and purchase cost, but also includes the additional welfare gained from product differentiation \cite{A-1}. To implement it, a transaction coefficient denoted as $\alpha_{ji}$ is introduced and imposed on the trade between consumer $j$ and producer $i$, and its detailed explanation is given later. Then, the welfare function of consumer $j$ is represented as
\begin{equation}
\begin{aligned}
\label{eq.wc}
WC_j &= U_j(y_j) - \sum_{i\in\mathcal{N}_j}\lambda_{ij}y_{ji}+\sum_{i\in\mathcal{N}_j}\alpha_{ji}y_{ji} \\
&= U_j(y_j) + \left(\alpha_j-\lambda_j\right)^T \mathbf{y}_j,\\
\text{with}&\quad y_j^{min}\le y_j\le y_j^{max}, \quad\forall j\in\mathcal{N}_C,
\end{aligned}
\end{equation}
where $\alpha_{j}=(\alpha_{ji})_{i\in\mathcal{N}_j}, \lambda_{j}=(\lambda_{ij})_{i\in\mathcal{N}_j}$ and $\mathbf{y}_{j}=(y_{ji})_{i\in\mathcal{N}_j}$. Note that in (4), a linear function with constant marginal benefit (i.e., $\alpha_{ji}y_{ji}$) is used to calculate the additional welfare brought by product differentiation \cite{31, A-1,38,42,43,33,51}. This is because the additional welfare arising from product differentiation generally have either constant marginal benefit or diminishing marginal benefit.

In the following, the composition of the transaction coefficients is explained, followed by two examples. In the P2P trading market, all transactions can be described by a set of criteria $\mathcal{S}$. The criterion can be the type of generation, the level of emission, the level of reliability, the location or rating of the producer, etc. Based on these criteria, the transaction coefficient $\alpha_{ji}$, used to reflect product differentiation, can be composed of two parts:  \textit{criterion values} and \textit{trade characteristics}. Specifically, under criterion $s\in\mathcal{S}$, the preference degree or relevant cost of consumer $j$ is characterized by the parameter $r_j^s$ (called the \textit{criterion value}), and consumer $j$'s objective evaluation of the trade with producer $i$ under criterion $s$ is expressed by parameter $d_{ji}^s$ (called the \textit{trade characteristic}). Then, the transaction coefficient $\alpha_{ji}$ between consumer $j$ and producer $i$ can be calculated as
\begin{equation}
\label{alpha}
\alpha_{ji} = \sum_{s\in\mathcal{S}}r_j^s d_{ji}^{s}.
\end{equation}
For instance, if the transaction coefficient is for the purpose of increasing local consumption, the `distance' can be used as a single criterion. Then, the \textit{criterion value} is the preference degree of consumer $j$ for local consumption with unit $\$\cdot \text{kWh}^{-1}\cdot  \text{km}^{-1}$, and the \textit{trade characteristic} $d_{ji}$ is the Euclidean distance between consumer $j$ and producer $i$ with unit $\text{km}$ \cite{A-1}. If the transaction coefficient is used to implement network fees for electricity network owners, the \textit{criterion value} is the network usage charge per unit electrical distance, and the \textit{trade characteristic} $d_{ji}$ is the electrical distance between consumer $j$ and producer $i$ \cite{33,38}.

\subsection{Network Constraints}
Since the energy transfers of P2P transactions are inseparable from the distribution network, the network constraints, including voltage and line flow constraints, should be considered to prevent overvoltage and congestion. In this paper, the voltage changes and line flows are calculated by using the voltage sensitivity coefficient and the power transfer distribution factor proposed in \cite{41}, respectively. The corresponding voltage and line flow constraints are formulated as follows
\begin{gather}
V_i^{min}\le V_i \le V_i^{max},\quad \forall i\in\mathcal{N}_B,\label{cons_volt}\\
-F_{\ell}^{max}\le F_{\ell} \le F_{\ell}^{max}, \quad\forall \ell\in\mathcal{N}_L,\label{cons_flow}
\end{gather}
where $V_i^{min}$ and $V_i^{max}$ are the lowest and highest voltage magnitudes, respectively. $F_{\ell}^{max}$ is the maximum capacity allowed for line $\ell$.

\subsection{Optimization Problem}
The problem of maximizing social welfare is widely considered \cite{ A-1, A-2, 38, 43,42}, where the social welfare consists of the welfare of all producers and consumers. Based on the welfare functions (\ref{eq.wp}) and (\ref{eq.wc}) and network constraints (\ref{cons_volt}) and (\ref{cons_flow}), the problem of maximizing social welfare is formulated as
\begin{subequations}
\label{eq.f}
\begin{equation}
\begin{aligned}
\mathop{\max}\limits_{\mathbf{x}, \mathbf{y}}\  \sum_{j\in\mathcal{N}_C}U_j(y_j)-\sum_{i\in\mathcal{N}_P}C_i(x_i)
+\sum_{j\in\mathcal{N}_C}\sum_{i\in\mathcal{N}_j}\alpha_{ji}y_{ji},\label{eq.fa}
\end{aligned}
\end{equation}
\begin{align}
\text{s.t.}\ &x_{ij} =y_{ji}, \quad\forall i\in\mathcal{N}_P, \ \forall j\in\mathcal{N}_i, \label{eq.fb} \\
&x_{ij} \ge 0, \quad\forall i\in\mathcal{N}_P, \ \forall j\in\mathcal{N}_i, \label{eq.fg}\\
&y_{ji} \ge 0, \quad\forall j\in\mathcal{N}_C, \ \forall i\in\mathcal{N}_j,\\
&x_i^{min}\le x_i=\sum\nolimits_{j\in\mathcal{N}_i}x_{ij} \le x_i^{max}, \quad\forall i\in\mathcal{N}_P,\label{eq.fe}\\
&y_j^{min}\le y_j=\sum\nolimits_{i\in\mathcal{N}_j}y_{ji}\le y_j^{max}, \quad\forall j\in\mathcal{N}_C,\label{eq.ff}\\
&V_b^{min}\le V_b \le V_b^{max},\quad \forall b\in\mathcal{N}_B,\label{eq_volt}\\
&-F_{\ell}^{max}\le F_{\ell} \le F_{\ell}^{max}, \quad\forall \ell\in\mathcal{N}_L,\label{eq_flow}
\end{align}
\end{subequations}
where $\mathbf{x}=(\mathbf{x}_{1}^T, \mathbf{x}_{2}^T, ..., \mathbf{x}_{n_p}^T)^T$ and $\mathbf{y}=(\mathbf{y}_{1}^T, \mathbf{y}_{2}^T, ..., \mathbf{y}_{n_c}^T)^T$ denote all transactions in the market.

\section{Market Trading Mechanism}
To solve the multi-bilateral economic dispatch problem (\ref{eq.f}), an accelerated distributed clearing algorithm and a selection strategy are proposed, which constitute our scalable energy trading mechanism.

\subsection{Problem Decoupling}
Before solving problem (13), the coupling constraint (\ref{eq.fb}) is dealt with first. Let $\lambda_{ij}$ denote the Lagrange multiplier to decouple (\ref{eq.fb}). Then, the Lagrange function associated to (\ref{eq.f}) denoted by $L(\cdot)$ is
\begin{equation}
\begin{aligned}
\label{eq.lf}
L(\mathbf{x}&,\mathbf{y}, \lambda)= \sum_{j\in\mathcal{N}_C}U_j ( y_j ) -\sum_{i\in\mathcal{N}_P}C_i ( x_i )\\
&+\sum_{j\in\mathcal{N}_C}\sum_{i\in\mathcal{N}_j}\alpha_{ji}y_{ji}+\sum_{i\in\mathcal{N}_P}\sum_{j\in\mathcal{N}_i}\lambda_{ij}\left( x_{ij}-y_{ji} \right),
\end{aligned}
\end{equation}
where $\lambda=(\lambda_{1}^T, \lambda_{2}^T, ..., \lambda_{n_p}^T)^T$ is a column vector composed of all Lagrange multipliers $\lambda_{ij}$. According to the Lagrange function (\ref{eq.lf}), the dual function $q(\lambda)$ is established as
\begin{equation}
\begin{aligned}
\label{eq.df}
q&(\lambda) =  \mathop{\max}\limits_{\mathbf{x}, \mathbf{y}}L(\mathbf{x},\mathbf{y}, \lambda)
\\
&=  \sum_{i\in\mathcal{N}_P}\underbrace{-C_i ( x_i ) +\lambda_{i}^T\mathbf{x}_{i}}_{q_i(\lambda_i)} + \sum_{j\in\mathcal{N}_C}\underbrace{U_j ( y_j ) + \left(\alpha_{j}-\lambda_{j}\right)^T\mathbf{y}_{j}}_{q_j(\lambda_j)},
\end{aligned}
\end{equation}
where each component of $\mathbf{x}$ and $\mathbf{y}$ is bounded by local constraints, i.e., $\mathbf{1}^T\mathbf{x}_{i}\in [x_i^{min}, x_i^{max}]$ and $\mathbf{1}^T\mathbf{y}_j\in [y_j^{min}, y_j^{max}]$. Then, the dual problem is as follows
\begin{equation}
\label{eq.d}
\mathop{\min}\limits_{\lambda}\ q(\lambda).
\end{equation}

Note that the dual problem has the same optimal solution as the original problem due to the strong duality, which can be proved using Proposition 5.2.1 in \cite{26}.


\subsection{Accelerated Market Clearing Algorithm}
To solve the dual problem (\ref{eq.d}), a distributed algorithm based on Nesterov's accelerated gradient is proposed, which is shown in Algorithm 1. 
\begin{algorithm}[h]  
  \caption{Accelerated Market Clearing Algorithm}
  \label{alg:Framwork}  
  \begin{algorithmic}[1]  
    \Require 
		\\For $i\in\mathcal{N}_P$, sets $\gamma^{1}=1$ and $\hat{\lambda}_{ij}^{1}=\lambda_{ij}^{0}, \forall j\in\mathcal{N}_i$.\\
		For $j\in\mathcal{N}_C$, receives $\hat{\lambda}_{ij}^{1}$ from all $i\in\mathcal{N}_j$.
    \Ensure \\
		Update $x_{ij}^{k}$ for $i\in\mathcal{N}_P$ and update $y_{ji}^{k}$ for $j\in\mathcal{N}_C$,
		\begin{equation}
		\begin{aligned}
		\label{eq.3.1}
		&x_{ij}^{k}=\mathop{\arg\max}\limits_{x_{ij}\ge 0}\left\{ -C_i(x_i)+\hat{\lambda}_{ij}^{k}x_{ij}\right\},\quad \forall j\in\mathcal{N}_i,\\
		&s.t.\ (\ref{eq.fe}),\ (\ref{eq_volt}),\ \text{and}\ (\ref{eq_flow}).
		\end{aligned}
		\end{equation}
		\begin{equation}
		\begin{aligned}
		\label{eq.3.2}
		&y_{ji}^{k}=\mathop{\arg\max}\limits_{y_{ji}\ge 0}\left\{ U_j(y_j)+\big(\alpha_{ji}-\hat{\lambda}_{ij}^{k}\big)y_{ji} \right\},\quad \forall i\in\mathcal{N}_j,\\
		&s.t.\ (\ref{eq.ff})\ \text{and}\ (\ref{eq_volt}).
		\end{aligned}
		\end{equation}
		\\
		Receive $y_{ji}^{k}$ from $j\in\mathcal{N}_i$ for $i\in\mathcal{N}_P$.\\
		Update $\lambda_{ij}^{k}$ for $i\in\mathcal{N}_P$,
		\begin{equation}
		\label{eq.3.3}
		\lambda_{ij}^{k}=\hat{\lambda}_{ij}^{k}-\eta_{ij}\left(x_{ij}^{k}-y_{ji}^{k}\right),\quad \forall j\in\mathcal{N}_i.
		\end{equation}
		\\
		Update $\gamma^{k+1}=\frac{\left(k+1\right)\left(1+\sqrt{1+4\left(\frac{\gamma^{k}}{k}\right)^2}\right)}{2}$ and $\hat{\lambda}_{ij}^{k+1}$ for $i\in\mathcal{N}_P, j\in\mathcal{N}_i$,
		\begin{equation}
		\label{eq.3.4}
\hat{\lambda}_{ij}^{k+1}=\lambda_{ij}^{k}+\frac{(k+1)(\gamma^{k}-k)}{k\gamma^{k+1}}\left(\lambda_{ij}^{k}-\lambda_{ij}^{k-1}\right).
		\end{equation}
\\
		Receive $\hat{\lambda}_{ij}^{k+1}$ from $i\in\mathcal{N}_j$ for $j\in\mathcal{N}_C$.
  \end{algorithmic}  
\end{algorithm}

In \textbf{Initialization}, each producer $i\in\mathcal{N}_P$ determines initial trading price $\lambda_{ij}^{0}$ with neighboring consumers and sets $\hat{\lambda}_{ij}^{1} = \lambda_{ij}^{0},\ \gamma^{1} =1$. Each consumer $j\in\mathcal{N}_C$ receives $\hat{\lambda}_{ij}^{1}$ from neighboring producers. In \textbf{Iteration}, each producer and consumer update the generation $x_{ij}^k$ and demand $y_{ji}^k$ with (\ref{eq.3.1}) and (\ref{eq.3.2}) to maximize their own welfare, respectively. Sub-problems (\ref{eq.3.1}) and (\ref{eq.3.2}) can be solved by using the Lagrange relaxation method in \cite{25}. The consumer's updated demand $y_{ji}^{k}$ is then sent to the corresponding producer. Based on the updated demand and generation, each producer updates the price $\lambda_{ij}^k$ to promote the social welfare according to (\ref{eq.3.3}) with step-size $\eta_{ij}$ in Step 5. Through the last term of (\ref{eq.3.3}), the energy consensus (i.e., $x_{ij}=y_{ji}$) is guaranteed. Then, the producer updates the parameter $\gamma^k$ and the acceleration term $\hat{\lambda}_{ij}^k$ using (\ref{eq.3.4}) in Step 6. Note that $\hat{\lambda}_{ij}$ is an interpolated point of $\lambda_{ij}$, which plays an acceleration role in the algorithm. Finally, the producer feeds back the latest price information $\hat{\lambda}_{ij}^{k+1}$ to the corresponding consumer. 
All the steps in Algorithm 1 can be implemented in a distributed manner without seeding any private information, such as $a_i$, $b_i$, $c_i$, $\omega_j$ and $\delta_j$. When the iteration triggers the stopping criteria, the algorithm stops running. The stopping criteria are set as follows
\begin{equation}
|x_{ij}^{k}-y_{ji}^{k}| \le\epsilon,\quad \forall i\in\mathcal{N}_P, j\in\mathcal{N}_i,
\end{equation}
where $\epsilon$ is extremely small positive numbers.

Unlike \cite{A-1}, our proposed algorithm does not need to consider the price consensus, since producer $i$ and consumer $j$ update $x_{ij}^k$ and $y_{ji}^k$ with the same price information $\hat{\lambda}_{ij}^k$. In this way, producers only need to send price information while consumers only need to send energy information, which reduces the exchanged information between players. In addition, compared with the primal-dual gradient method used in \cite{A-2}, our algorithm adds Steps 5 and 6, and replaces $\lambda_{ij}$ with $\hat{\lambda}_{ij}$ in Steps 3 and 7, which can improve the convergence rate. This fact will be verified in Theorem 1 and Corollary 1 of Section IV, and explained in Remark \ref{remark_rate}.

\subsection{Selection Strategy}
It can be seen from Algorithm 1 that the amount of computation and communication for each player largely depends on the number of neighbors (i.e., the number of trading partners). As a consequence, reducing the number of trading partners can reduce the amount of computation and communication. However, this may lose some social welfare due to artificially limiting possible transactions. In fact, there is a trade-off between the number of trading partners and optimal social welfare \cite{13}. Our goal is to develop a simple selection strategy that reduces the number of trading partners per player, while keeping the optimal social welfare as much as possible. 

Our selection strategy is based on a basic microeconomics principle: in maximizing social welfare, an allocation is ineffective if a good is not consumed by the buyers who value it most highly \cite[Chapter 7]{39}. Thus, if these ineffective transactions can be blocked before executing the clearing algorithm, we can achieve our goal.
One feasible strategy is that each consumer selects its preferred producers from neighbors and only choose them as its trading partners. In this way, not only the number of trading partners per player is reduced, but also each producer's energy can be allocated to the consumers who prefer it (i.e., efficient transactions in the market is preserved).

The key for the selection strategy is to find the preferred producers for each consumer. In problem (\ref{eq.f}), the transaction coefficient $\alpha_{ji}$ for product differentiation essentially reflects the preference of consumer $j$ to producer $i$. Hence, based on each consumer's transaction coefficients for neighboring producers, the following selection strategy is proposed with only two simple steps
\begin{enumerate}[\textbf{Step (1):}]
\item [\textbf{Step 1:}]
Each consumer $j\in\mathcal{N}_C$ normalizes transaction coefficients $\alpha_{ji}$ for each neighboring producers $i\in\mathcal{N}_j$ to [-1,1] and sets 0 as the preferred benchmark.
\item [\textbf{Step 2:}]
Each consumer $j\in\mathcal{N}_C$ only chooses neighboring producers whose normalized transaction coefficient is greater than or equal to 0 as its trading partners.
\end{enumerate}
In Step 1, the purpose of normalization is to explicitly reflect each consumer's preferred producers relatively. Then in Step 2 consumers choose these preferred producers as their trading partners instead of all neighboring producers. On one hand, this reduces the number of trading partners per player and thus reduces the amount of computation and communication per player. On the other hand, choosing the preferred producer preserves efficient transactions in the market and thus keeps the optimal social welfare as much as possible, which is illustrated in Section V-C.
\begin{remark}
The selection strategy is executed before the clearing algorithm starts. So it can be extendable to any clearing methods for P2P electricity markets. In addition, it neither has any computation burden nor brings any privacy issues since the transaction coefficients $\alpha_{ji}, i\in\mathcal{N}_j$ are the local information of consumer $j$.

\end{remark}

\section{Convergence Analysis}
This section analyzes the convergence rates of Algorithm 1 and the algorithm in \cite{A-2}. We first give the result of Algorithm 1 in Theorem 1.
\begin{theorem}\label{thm1}
Let the sequence $\lambda_{ij}^{k}$ be generated by Algorithm 1 and the step-size is set to $\eta_{ij}\in(0,1/L_{ij}]$, where $L_{ij} = \frac{\sigma_i+\sigma_j}{\sigma_i\sigma_j}$ and $\sigma_i$, $\sigma_j$ are the strong concave constants of the cost and utility function of producer $i$ and consumer $j$, respectively. Then, for any $k\ge1$
\begin{equation}
\label{pro1}
q(\lambda^k) - q(\lambda^*)
\le
\sum_{i\in\mathcal{N}_P}\sum_{j\in\mathcal{N}_i} \frac{2\left(\lambda_{ij}^0-\lambda_{ij}^*\right)^2}{\eta_{ij}k^2},
\end{equation}
where $\lambda^*$ is the optimal solution of the dual problem (\ref{eq.d}).
\end{theorem} 
\begin{IEEEproof}
The IEEEproof can be found in Appendix \ref{appendix_IEEEproof_theorem1}.
\end{IEEEproof}

Theorem 1 shows that the dual sequence of Algorithm 1 converges to the optimal value with the rate of $O(1/k^2)$. In the following, the convergence result is given for our proposed algorithm without acceleration term, which is consistent with the primal-dual gradient method in \cite{A-2}.

\begin{corollary}
The acceleration terms in Algorithm 1 are not considered, i.e., Steps 5 and 6 are ignored, and $\hat{\lambda}_{ij}^{k}$ in Steps 3 and 7 are replaced by $\lambda_{ij}^{k}$. Let $\eta_{ij}\in(0,1/L_{ij}]$, where $L_{ij}$ is the same as in Theorem 1. Then, for any $k \ge 1$
\begin{equation}
\label{corollary}
q(\lambda^k) - q(\lambda^*)
\le
\sum_{i\in\mathcal{N}_P}\sum_{j\in\mathcal{N}_i}
\frac{\left(\lambda_{ij}^0-\lambda_{ij}^*\right)^2}{2\eta_{ij}k}.
\end{equation}
\end{corollary}

\begin{IEEEproof}
The IEEEproof can be found in Appendix \ref{appendix_IEEEproof_corollary1}.
\end {IEEEproof}

\begin{remark}\label{remark_rate}
From Theorem 1 and Corollary 1, it can be concluded that our proposed algorithm has a faster convergence rate in the sense that the acceleration term $\hat{\lambda}_{ij}$ in Algorithm 1 can indeed improve the convergence rate.
\end{remark}

\section{Numerical simulations}
In this section, we first demonstrate the feasibility of our proposed clearing algorithm in IEEE 15-bus system. Then, we illustrate the effectiveness of the selection strategy through a Monte Carlo simulation experiment. Finally, we validate the scalability of our proposed mechanism compared with other state-of-the-art methods in a large-scale market.

\subsection{Simulation Setup}
The setup considers a standard IEEE 15-bus system with 14 prosumers (including 7 producers and 7 consumers), where producers are located at buses 1, 3, 4, 5, 9, 10, 11 and consumers are positioned in buses 2, 6, 7, 8, 12, 13, 14 in sequence. The bus 0 is the slack bus. All bus voltages are initialized to 1 p.u., limited to $[0.9, 1.1]$. The maximum flow of lines is 60kW. The producers and consumers parameters are taken from \cite{22}.
The transaction coefficients between consumers and producers are randomly generated from $[0,1)$. In addition, the step-size is set to $\eta_{ij}=0.1, i\in\mathcal{N}_P, j\in\mathcal{N}_i$ and the stopping criteria are set to 0.001.


%
\begin{table}[htb]
\renewcommand{\arraystretch}{1.3}
\caption{Energy traded between consumers and producers, unit kWh}
\label{table_trading}
\centering
\resizebox{\columnwidth}{!}{
\begin{tabular}{c|cccccccc}
\hline
&Pro.1&Pro.2&Pro.3&Pro.4&Pro.5&Pro.6&Pro.7&Sum\\
\hline
Con.1&0.00& 	26.28& 	3.44& 	0.00& 	0.00& 	5.43& 	0.00& 	35.15\\ 
Con.2&9.82& 	16.84& 	0.00& 	0.00& 	0.00& 	0.00& 	0.00& 	26.66\\ 
Con.3&0.00&	0.00& 	0.00& 	0.00& 	15.85& 	14.41& 	20.56& 	50.82\\  
Con.4&4.69& 	0.00& 	1.98& 	0.00& 	0.00& 	0.00& 	23.33& 	30.00\\  
Con.5&5.85& 	0.00& 	0.00& 	25.47& 	10.60& 	19.97& 	0.00& 	61.89\\  
Con.6&0.00& 	0.00& 	0.00& 	0.00& 	7.50& 	0.00& 	0.00& 	7.50\\  
Con.7&18.75& 	0.00& 	12.58& 	0.00& 	0.00& 	0.00& 	0.00& 	31.33\\  
Sum&39.11& 	43.12& 18.00& 	25.47& 	33.95& 	39.81& 	43.88& 	243.34\\  
\hline
\end{tabular}}
\end{table}

\begin{figure}[htb]
\centering
\subfloat[]{\includegraphics[width=1.66in,height=1.2in]{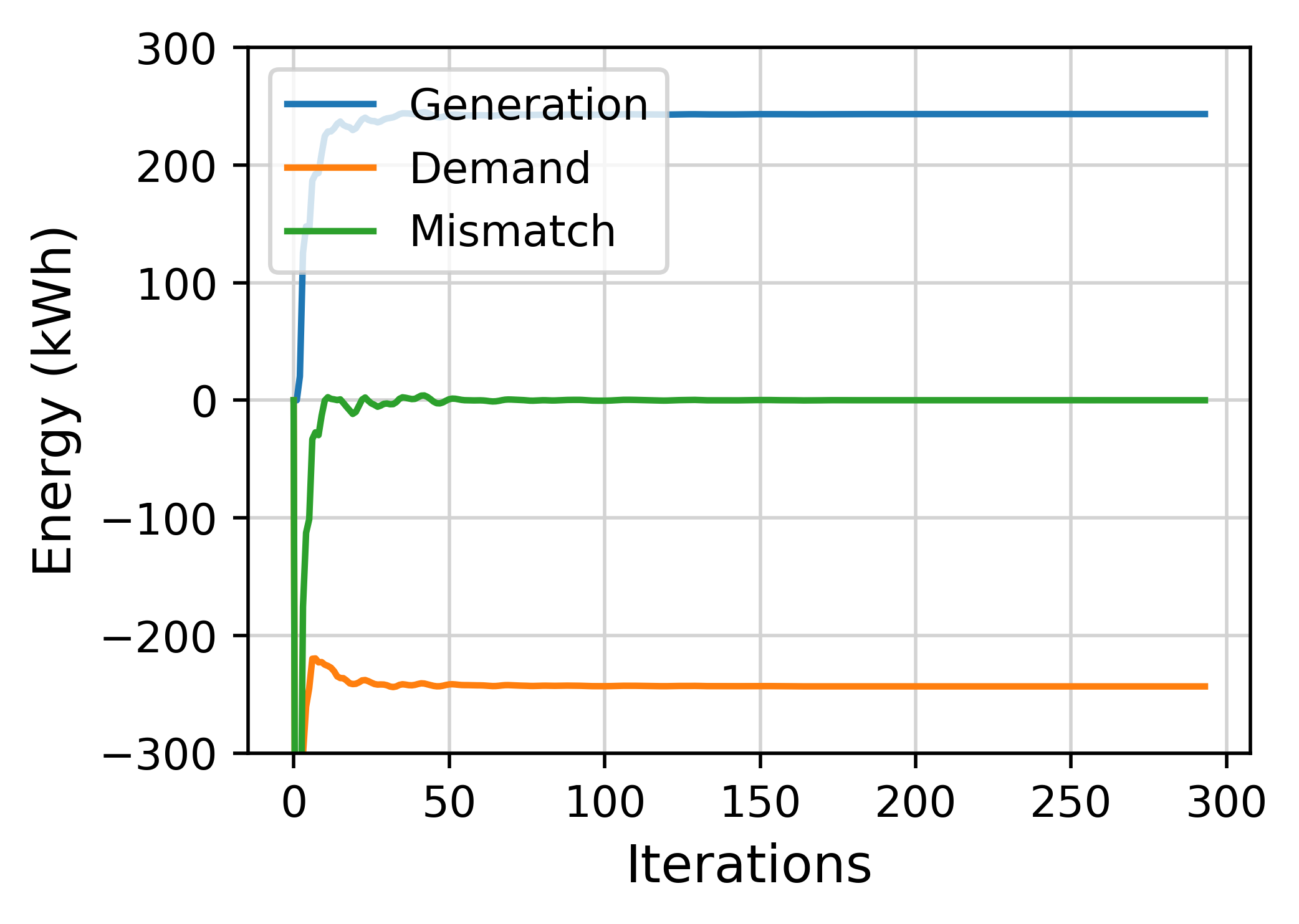} \label{fig_1a}}
\hfil
\subfloat[]{\includegraphics[width=1.66in,height=1.2in]{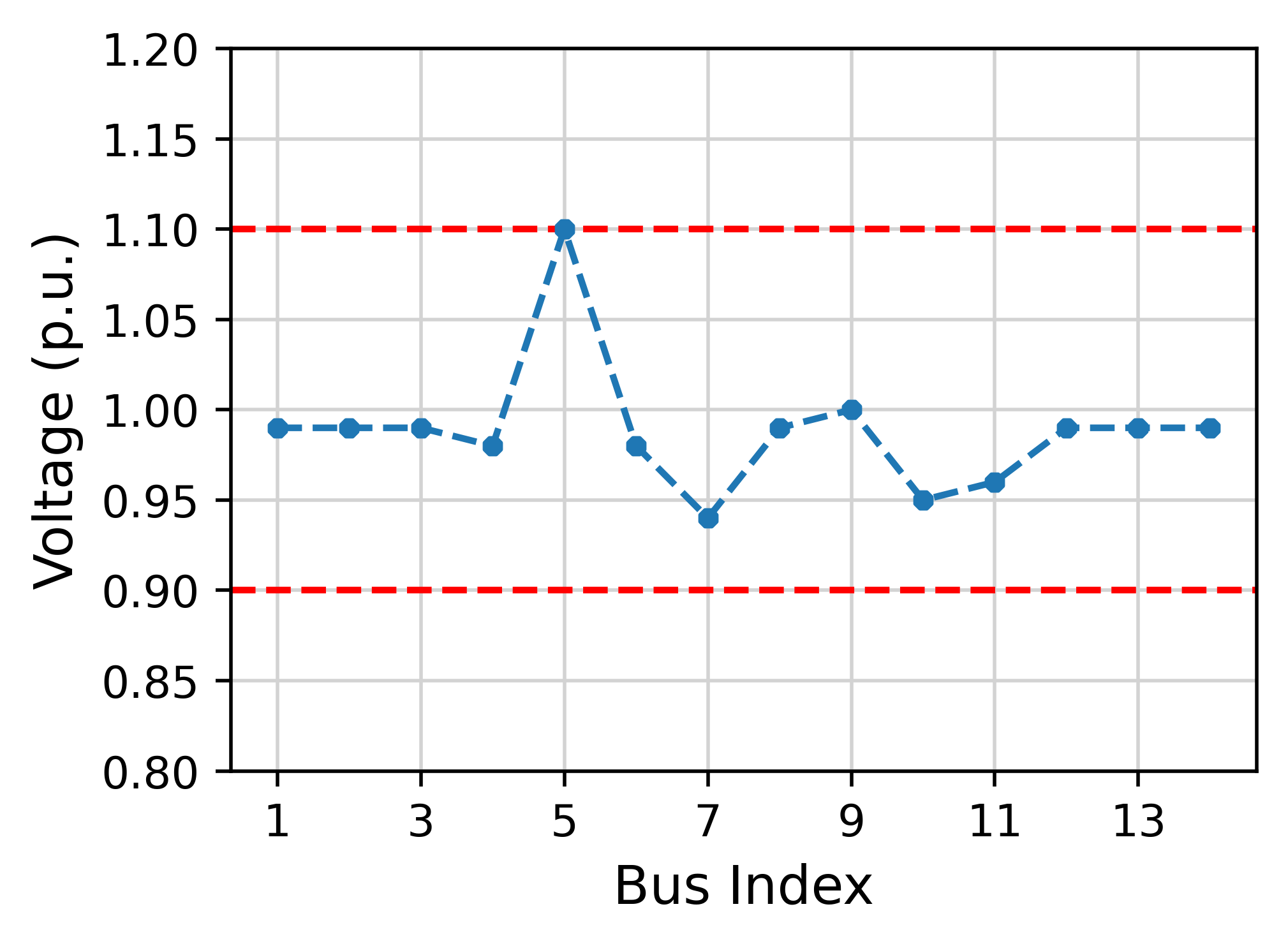} \label{fig_1b}}
\hfil
\subfloat[]{\includegraphics[width=1.66in,height=1.2in]{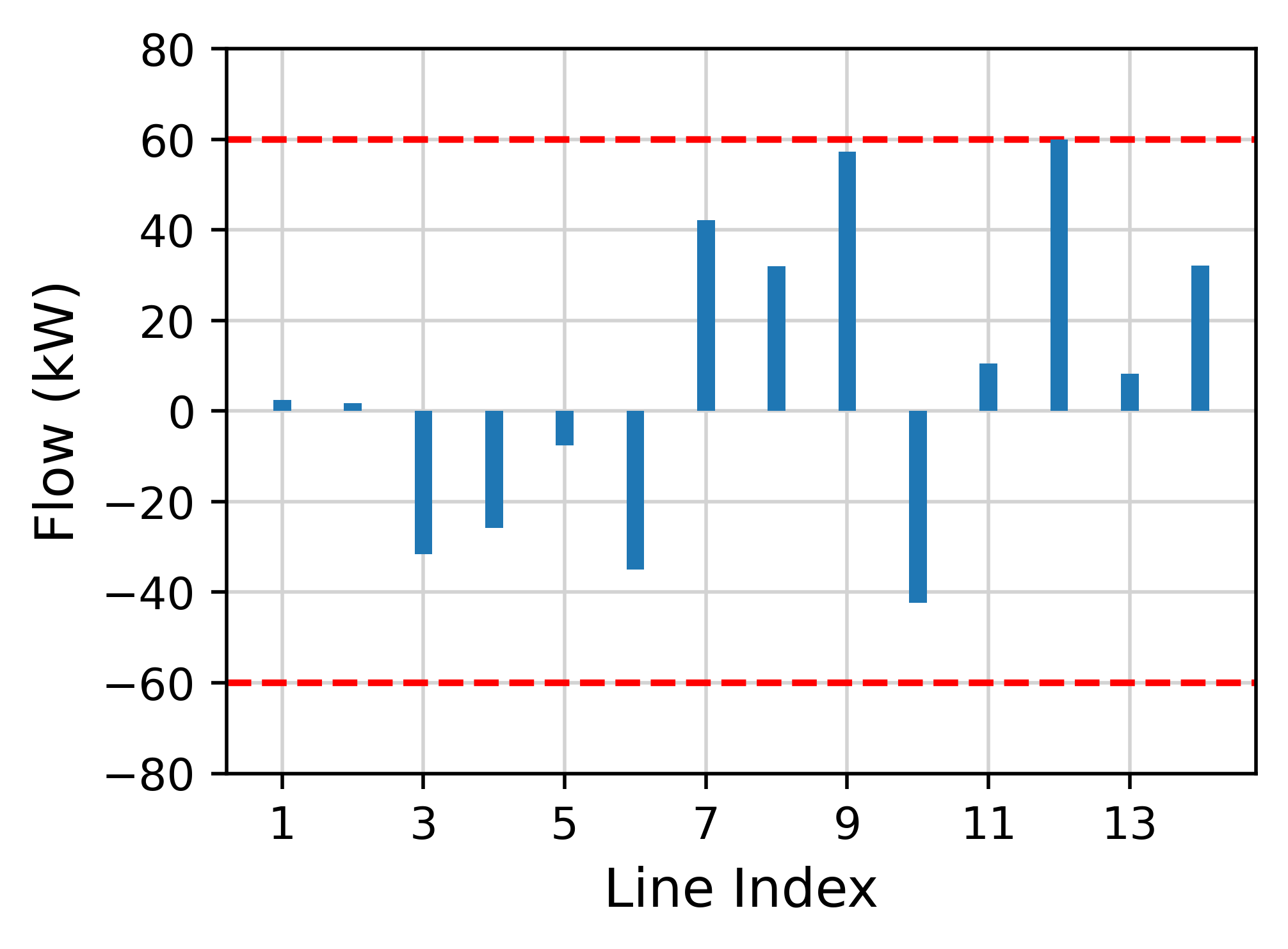}\label{fig_1c}}
\hfil
\subfloat[]{\includegraphics[width=1.66in,height=1.2in]{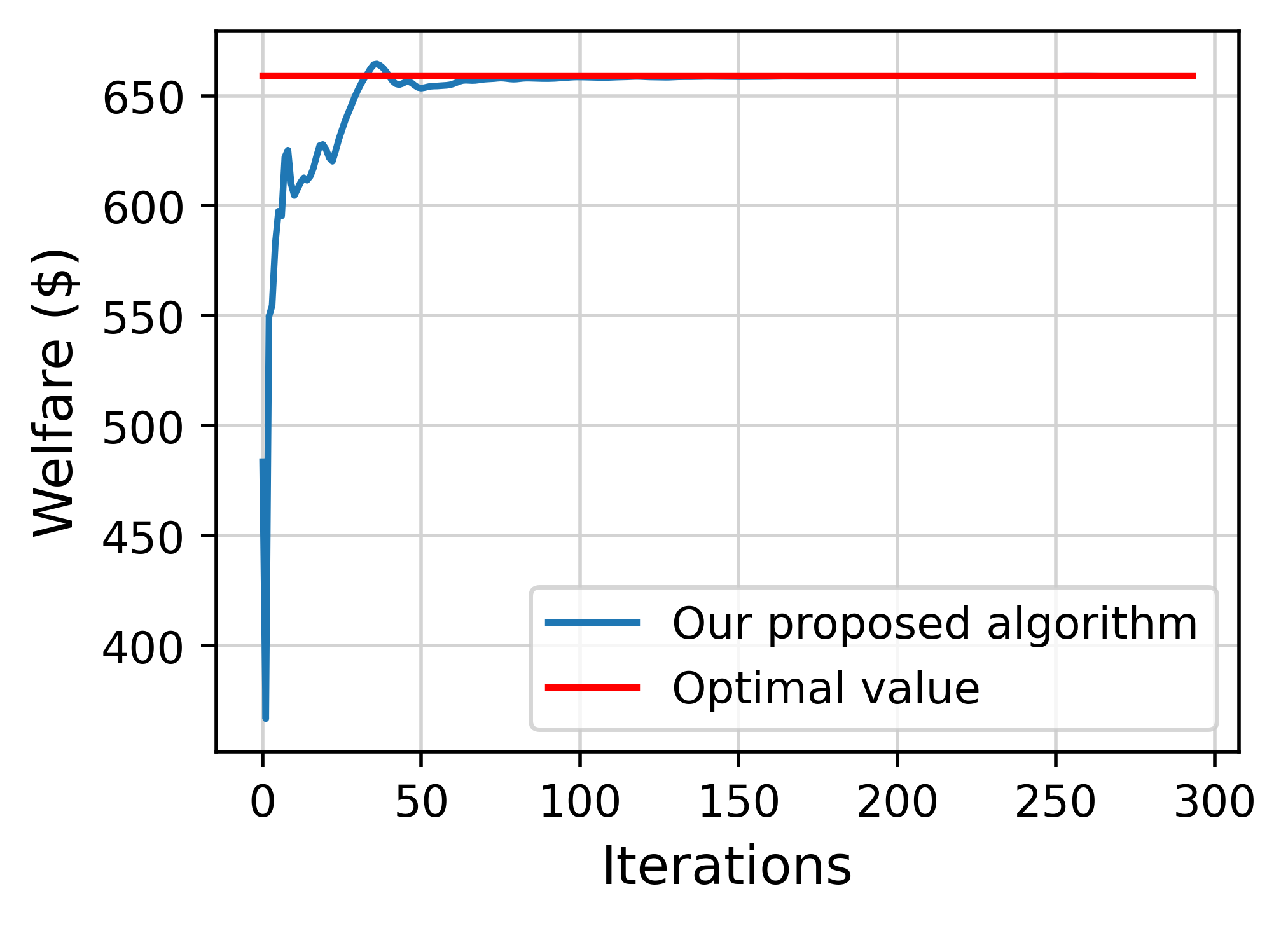} \label{fig_1d}}
\caption{Feasibility analysis of clearing algorithm: (a) The balance of total generation and total demand, (b) Bus voltages, (c) Line flows, (b) The evolution of social welfare.}
\label{fig_1}
\end{figure}
\subsection{Feasibility of Clearing Algorithm}
In this subsection, we illustrate the feasibility of our proposed clearing algorithm as indicated in Table \ref{table_trading} and Fig. \ref{fig_1}. Table \ref{table_trading} shows energy traded between consumers and producers.
Fig. \ref{fig_1}(a) shows that total generation and demand are balanced. Fig. \ref{fig_1}(b) and (c) show that energy transactions do not violate network constraints. Fig. \ref{fig_1}(d) illustrates that social welfare generated by our proposed algorithm can asymptotically reach an optimal value with iterations. Therefore, we conclude that our proposed algorithm can converge to an optimal solution with energy and network constraints.

In addition, Table \ref{table_trading} shows that consumers only conduct efficient energy transactions with some producers. For example, Consumer 1 only buys energy from Producers 2, 3, and 6. This is due to Consumer 1 has a larger transaction coefficient with Producers 2, 3, and 6 ($\alpha_{11}, \alpha_{12}, \alpha_{13}, \alpha_{14}, \alpha_{15}, \alpha_{16}, \alpha_{17}: 0.54, 0.71, 0.60, 0.54, 0.42, 0.64, 0.43$). It implies that consumers are more willing to enter into transactions with their preferred producers.
\begin{figure}[!h]
\centering
\subfloat[]{\includegraphics[width=3.2in,height=2.2in]{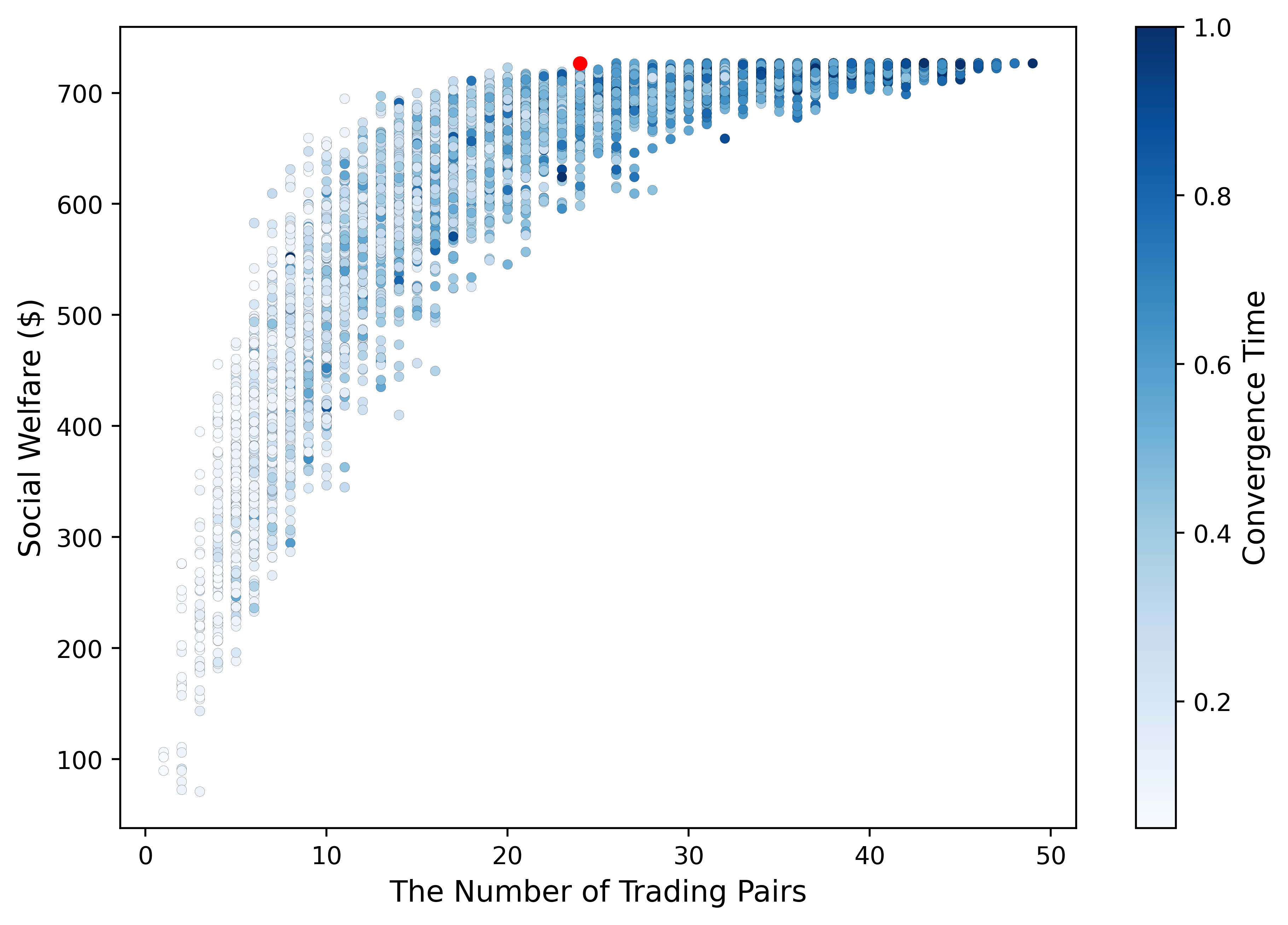}
\label{Fig_select_strategy}}
\hfil
\subfloat[]{\includegraphics[width=3.2in,height=1.2in]{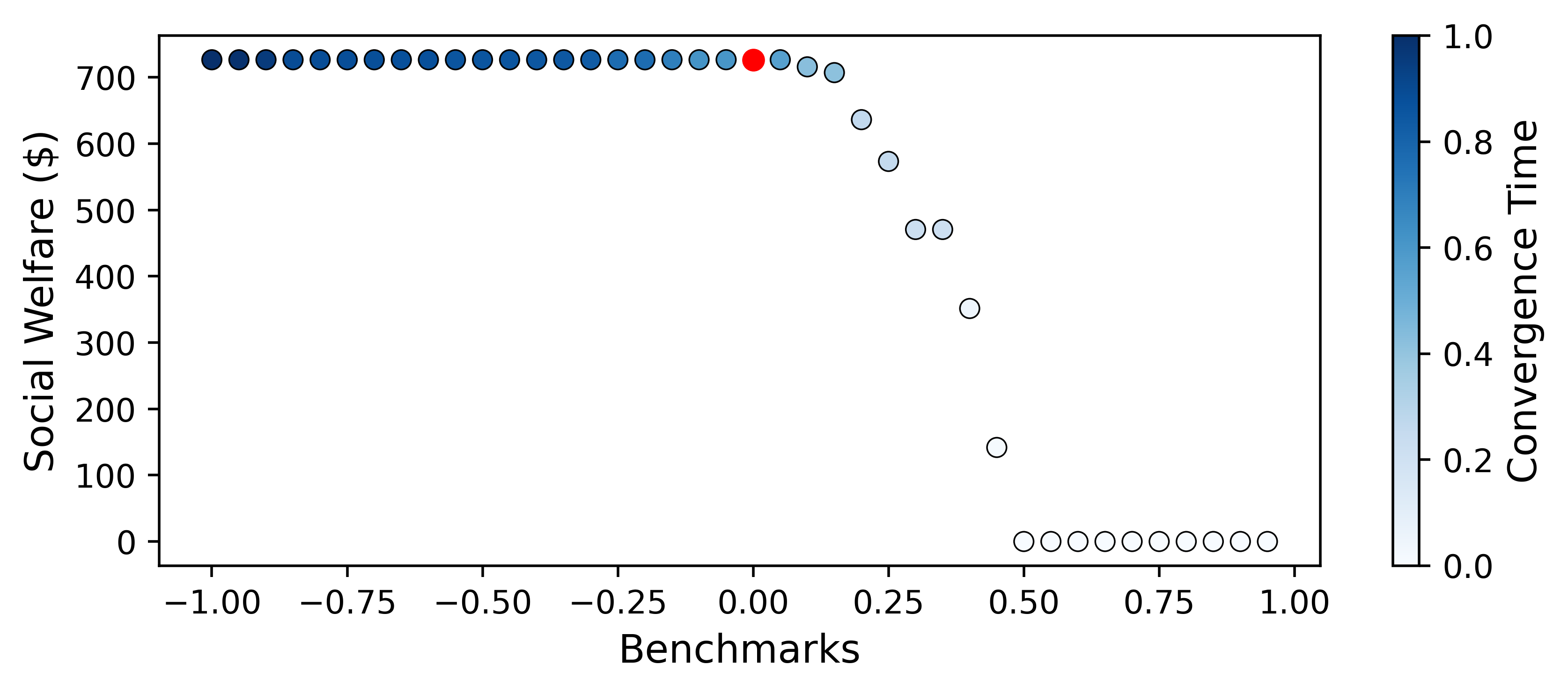}
\label{Fig_select_benchmark}}
\hfil
\caption{Effectiveness of the selection strategy: (a) Number of trading pairs, social welfare, and convergence time for 7000 simulations, (b) Social welfare and convergence time under different benchmark values.}
\label{Fig_SS}
\end{figure}
\subsection{Effectiveness of Selection Strategy}
To verify the effectiveness of the proposed selection strategy, we conduct a Monte Carlo experiment in which consumers randomly select different trading partners in each simulation. We perform 7000 simulations and record the number of trading pairs (i.e., sum of the number of trading partners of all consumers), social welfare and convergence time in each simulation, as indicated in Fig. \ref{Fig_SS}(a). It shows that as the number of trading pairs reduces, both social welfare and convergence time are trending down. This is because reducing the number of trading pairs can reduce the amount of computation and communication, at the expense of some social welfare. Fig. \ref{Fig_SS}(a) also shows that a reasonable reduction of trading pairs can not only decrease convergence time but also hardly lose social welfare, as shown by the red dot, which is the case of the proposed selection strategy.
In the next subsection, we further illustrate the performance of the selection strategy through specific numerical comparisons.

In the selection strategy, we set the normalized transaction coefficient of 0 as the preferred benchmark. However, is the benchmark set reasonable? We conduct several simulations under different benchmark values, as indicted in Fig. \ref{Fig_SS}(b). It shows that with the increase of benchmark value, convergence time gradually decreases while social welfare is almost unchanged until the benchmark value reaches 0.15. This is because as the benchmark value increases, the number of trading partners selected by consumers decreases, which can reduce the amount of computation and communication. In addition, since consumers can still conduct efficient transactions with their preferred producers, the loss of social welfare is limited. When consumers cannot trade with enough producers, the loss of social welfare is exacerbated. Note that the red dot in Fig. \ref{Fig_SS}(b) is the case of the benchmark value we set, which is appropriate in the sense that it can effectively reduce convergence time without causing a significant reduction in social welfare.

\begin{table}[!t]
\renewcommand{\arraystretch}{1.3}
\caption{Performance of our proposed mechanism and other methods}
\label{table_scalability}
\centering
\resizebox{\columnwidth}{!}{
\begin{threeparttable}
\begin{tabular}{ccccc}
\hline
\bfseries & \bfseries M1 & \bfseries M2 & \bfseries M3 & \bfseries M4\\
\hline
No. of Iteration &5464&4954&3904&3352\\
Convergence Time (s) &5520.95&4034.57&3160.67&1700.13 \\
Social Welfare (\$) &28115.21&28114.10&28113.70&28098.34\\
\hline
\end{tabular}
\end{threeparttable}
}
\end{table}

\subsection{Scalability of the Algorithm}
In this subsection, we illustrate the scalability of our proposed mechanism by comparing with other methods in a large-scale P2P market. We expand the original 14-prosumers market to a 500-prosumers market. These methods for comparison are: M1: the RCI method reported in \cite{A-1}, M2: the DBET method proposed in \cite{A-2}, M3: our proposed algorithm without selection strategy, M4: our proposed algorithm with selection strategy. For each method, we set the same stopping criteria and run multiple simulations to determine its optimal step-size. We record the number of iterations, convergence time, and social welfare for each method, as indicated in Table \ref{table_scalability}. It can be found that M3 and M4 require fewer iterations than M1 and M2, which is due to the faster convergence rate. The convergence time of M4 is less, because the selection strategy can further reduce the amount of computation and communication. Less convergence time means better scalability for the electricity market.

\section{CONCLUSION}
In this paper, we consider the multi-bilateral economic dispatch problem with product differentiation. To meet the scalability challenge of the problem, we propose a scalable mechanism including an accelerated distributed clearing algorithm and a novel selection strategy. The accelerated distributed algorithm using Nesterov's accelerated gradient has a faster convergence rate than many existing methods. The selection strategy based on consumer preferences can reduce the amount of computation and communication per player by reducing the number of inefficient trading partners for consumers, while keeping the optimal social welfare as much as possible.


There are many interesting problems for future research, for instance, incorporating energy storage devices, employing more general cost and utility functions for prosumers, and protecting prosumers' privacy.



%

\appendices
\section{IEEEproof of Theorem 1}\label{appendix_IEEEproof_theorem1}
We first give five lemmas for proving Theorem 1.
\begin{lemma} \label{lemma1}
The local dual functions $q_i(\lambda_i)$ and $q_j(\lambda_j)$ defined in (\ref{eq.df}) have Lipschitz continuous gradient with Lipschitz constants $\frac{1}{\sigma_i}$ and $\frac{1}{\sigma_j}$, that is, $\|\nabla q_i(\mu_i)-\nabla q_i(\lambda_i)\| \le \frac{1}{\sigma_i}\|\mu_i-\lambda_i\|$ and $\|\nabla q_j(\mu_j)-\nabla q_j(\lambda_j)\| \le \frac{1}{\sigma_j}\|\mu_j-\lambda_j\|$.
\end{lemma}
\begin{IEEEproof}
According to the definition of $q_i(\lambda_i)$ in (\ref{eq.df}), let $
\mathbf{x}_i(\lambda_i)=\mathop{\text{arg max}}\limits_{\mathbf{x}_{i}}\left(-C_i(x_i)+\lambda_{i}^T\mathbf{x}_{i}\right)$ and $
\mathbf{x}_i(\mu_i)=\mathop{\text{arg max}}\limits_{\mathbf{x}_{i}}\left(-C_i(x_i)+\mu_{i}^T\mathbf{x}_{i}\right)$.
By using the optimality conditions in \cite{32}, we have
		\begin{align*}
\Big\langle -\nabla C_i(\mathbf{x}_i(\lambda_i))+\lambda_i,\mathbf{x}_i(\mu_i) -\mathbf{x}_i(\lambda_i) \Big\rangle &\le 0,\\
\Big\langle -\nabla C_i(\mathbf{x}_i(\mu_i))+\mu_i, \mathbf{x}_i(\lambda_i) - \mathbf{x}_i(\mu_i) \Big\rangle &\le 0.
		\end{align*}
Then, adding the above two inequalities, it gives
		\begin{align*}
&\Big\langle -\nabla C_i(\mathbf{x}_i(\lambda_i))+\nabla C_i(\mathbf{x}_i(\mu_i)), \mathbf{x}_i(\mu_i) -\mathbf{x}_i(\lambda_i)\Big\rangle\\
&\le \| \mu_i-\lambda_i\| \|\mathbf{x}_i(\mu_i) - \mathbf{x}_i(\lambda_i)\|,
		\end{align*}
It follows from the strong concavity of $-C_i(\cdot)$ that
		\begin{align*}
\sigma_i\|\mathbf{x}_i(\mu_i) - \mathbf{x}_i(\lambda_i)\|^2 &\le \Big\langle \mu_i-\lambda_i, \mathbf{x}_i(\mu_i) - \mathbf{x}_i(\lambda_i)\Big\rangle,
		\end{align*}
where $\sigma_i$ is the strong concave constant. Then, the Lipschitz continuous gradient of $q_i(\lambda_i)$ follows from the fact that $\nabla q_i(\mu_i)-\nabla q_i(\lambda_i)=\mathbf{x}_i(\mu_i) - \mathbf{x}_i(\lambda_i)$.
Similarly, $q_j(\lambda_j)$ also has Lipschitz continuous gradient.
\end{IEEEproof}

\begin{lemma} \label{lemma2}
For any $\lambda, \mu\in\mathbb{R}^{n_p n_c}$, the following inequality holds
\begin{equation*}
\label{ieq.a0}
q(\lambda)\le q(\mu)+\langle \lambda-\mu,\nabla q(\mu) \rangle + \sum_{i\in\mathcal{N}_P}\sum_{j\in\mathcal{N}_i} \frac{L_{ij}}{2}\left(\lambda_{ij}-\mu_{ij}\right)^2,
\end{equation*}
where $L_{ij} = \frac{\sigma_i+\sigma_j}{\sigma_i\sigma_j}$.
\end{lemma}
\begin{IEEEproof}
According to Lemma 1, we have
		\begin{equation*}
		q_i(\lambda_i)\le q_i(\mu_i)+\langle \lambda_i-\mu_i,\nabla q_i(\mu_i) \rangle + \sum_{j\in\mathcal{N}_i}\frac{1}{2\sigma_i}\left(\lambda_{ij}-\mu_{ij}\right)^2.
		\end{equation*}
Define $q_x(\lambda)=\sum_{i\in\mathcal{N}_P}q_i(\lambda_i)$, then summing all $q_i(\lambda_i)$ for all $i\in\mathcal{N}_P $ yields
		\begin{equation}
		\label{ieq.a1}
		q_x(\lambda)\le q_x(\mu)+\langle \lambda-\mu,\nabla q_x(\mu) \rangle + \sum_{i\in\mathcal{N}_P}\sum_{j\in\mathcal{N}_i} \frac{1}{2\sigma_i}\left(\lambda_{ij}-\mu_{ij}\right)^2.
		\end{equation}
Similarly, we can show that
		\begin{equation}
		\label{ieq.5.2.3}
		q_y(\lambda)\le q_y(\mu)+\langle \lambda-\mu,\nabla q_y(\mu) \rangle + \sum_{i\in\mathcal{N}_P}\sum_{j\in\mathcal{N}_i} \frac{1}{2\sigma_j}\left(\lambda_{ij}-\mu_{ij}\right)^2,
		\end{equation}
where $q_y(\lambda)=\sum_{j\in\mathcal{N}_C}q_j(\lambda_j)$. Then, combining (\ref{ieq.a1}) and (\ref{ieq.5.2.3}), we can draw the conclusion.
\end{IEEEproof}

\begin{lemma} \label{lemma3}
Let the sequence ${\gamma^{k}, \lambda_{ij}^{k}}$ be defined in Algorithm 1 with $\eta_{ij}\in(0,1/L_{ij}]$, where $L_{ij} = \frac{\sigma_i+\sigma_j}{\sigma_i\sigma_j}$. Then, the following inequality holds
\begin{align*}
&\frac{(\gamma^k)^2}{k^2} h^k-\frac{(\gamma^{k+1})^2}{(k+1)^2}h^{k+1}\\ 
\ge &\sum_{i\in\mathcal{N}_P}\sum_{j\in\mathcal{N}_i}\frac{1}{2\eta_{ij}}\left(u_{ij}^{k+1}\right)^2
-
\sum_{i\in\mathcal{N}_P}\sum_{j\in\mathcal{N}_i}\frac{1}{2\eta_{ij}}\left(u_{ij}^{k}\right)^2,
\end{align*}
where $h^k = q(\lambda^k)-q(\lambda^*)$, $u_{ij}^k = \frac{\gamma^{k}}{k}\lambda_{ij}^{k}-(\frac{\gamma^{k}}{k}-1)\lambda_{ij}^{k-1}-\lambda_{ij}^{*}$.
\end{lemma}
\begin{IEEEproof}
Define the quadratic approximation function $Q(\lambda, \mu)$ of $q(\lambda)$ as
\begin{equation*}
Q(\lambda, \mu) = q(\mu)+\langle \lambda-\mu,\nabla q(\mu) \rangle + \sum_{i\in\mathcal{N}_P}\sum_{j\in\mathcal{N}_i} \frac{1}{2\eta_{ij}}\left(\lambda_{ij}-\mu_{ij}\right)^2.
\end{equation*}
In light of Lemma 2, it can be obtained that
		\begin{equation}
		\label{ieq.qQ}
		q(\lambda) \le Q(\lambda, \mu).
		\end{equation}
Substituting $\lambda=\phi, \phi\in\mathbb{R}^{n_p n_c}$ and $\mu=\hat{\lambda}$ into $Q(\lambda, \mu)$, it has
		\begin{equation*}
		Q(\phi, \hat{\lambda}) = q(\hat{\lambda})+\langle \phi-\hat{\lambda},\nabla q(\hat{\lambda}) \rangle + \sum_{i\in\mathcal{N}_P}\sum_{j\in\mathcal{N}_i} \frac{1}{2\eta_{ij}}\left(\phi_{ij}-\hat{\lambda}_{ij}\right)^2,
		\end{equation*}
where $\phi:=\text{arg min}\{Q(\lambda,\hat{\lambda})\}$. In accordance with the first order optimality conditions, it obtains $\phi_{ij}=\hat{\lambda}_{ij}-\eta_{ij}\nabla q(\hat{\lambda}_{ij})$. Combining the convexity of $q(\lambda)$ (i.e., $q(\lambda)\ge q(\hat{\lambda})+\langle \lambda-\hat{\lambda},\nabla q(\hat{\lambda}) \rangle$) and $(\ref{ieq.qQ})$, we have
		\begin{equation}
		\begin{aligned}
		\label{ieq.a3}
		q(\lambda) - q(\phi) \ge& q(\lambda) - Q(\phi,\hat{\lambda})
		\\
		\ge& \left \langle \lambda - \hat{\lambda},\nabla q(\hat{\lambda}) \right\rangle - 
		\left \langle \phi-\hat{\lambda},\nabla q(\hat{\lambda})\right\rangle \\ 
		&-\sum_{i\in\mathcal{N}_P}\sum_{j\in\mathcal{N}_i} \frac{1}{2\eta_{ij}}\left(\phi_{ij}-\hat{\lambda}_{ij}\right)^2
		\\
		=&
		\sum_{i\in\mathcal{N}_P}\sum_{j\in\mathcal{N}_i} \frac{1}{2\eta_{ij}}\left(\phi_{ij}-\hat{\lambda}_{ij}\right)^2\\
		&+
		\sum_{i\in\mathcal{N}_P}\sum_{j\in\mathcal{N}_i} \frac{1}{\eta_{ij}}( \phi_{ij}-\hat{\lambda}_{ij})(\hat{\lambda}_{ij}-\lambda_{ij}),
		\end{aligned}
		\end{equation}
where the first equality uses $\nabla q(\hat{\lambda}_{ij}) =\frac{1}{\eta_{ij}} (\hat{\lambda}_{ij} - \phi_{ij})$. By using (\ref{ieq.a3}) with the points $(\lambda:=\lambda^k,\hat{\lambda}:=\hat{\lambda}^{k+1})$ and $(\lambda:=\lambda^*, \hat{\lambda}:=\hat{\lambda}^{k+1})$, it yields
		\begin{equation}
		\begin{aligned}
		\label{ieq.a4}
		h^k-&h^{k+1} \ge \sum_{i\in\mathcal{N}_P}\sum_{j\in\mathcal{N}_i}\frac{1}{2\eta_{ij}}\left(\lambda^{k+1}_{ij}-\hat{\lambda}^{k+1}_{ij}\right)^2\\ 
		&+ 
		\sum_{i\in\mathcal{N}_P}\sum_{j\in\mathcal{N}_i}\frac{1}{\eta_{ij}}( \lambda^{k+1}_{ij}-\hat{\lambda}_{ij}^{k+1})(\hat{\lambda}_{ij}^{k+1}-\lambda_{ij}^{k}),
		\end{aligned}
		\end{equation}
		\begin{equation}
		\begin{aligned}
		\label{ieq.a5}
		-h^{k+1}& \ge \sum_{i\in\mathcal{N}_P}\sum_{j\in\mathcal{N}_i}\frac{1}{2\eta_{ij}}\left(\lambda^{k+1}_{ij}-\hat{\lambda}^{k+1}_{ij}\right)^2 \\
		&+
		\sum_{i\in\mathcal{N}_P}\sum_{j\in\mathcal{N}_i}\frac{1}{\eta_{ij}}( \lambda^{k+1}_{ij}-\hat{\lambda}_{ij}^{k+1})(\hat{\lambda}_{ij}^{k+1}-\lambda_{ij}^{*}),
		\end{aligned}
		\end{equation}
where $h^k = q(\lambda^k) - q(\lambda^*)$. To get the relationship between $h^k$ and $h^{k+1}$, multiply the inequality $(\ref{ieq.a4})$ by $(\frac{\gamma^{k+1}}{k+1}-1)$ and add it to $(\ref{ieq.a5})$.
Then, multiplying both sides of the obtained result by $\frac{\gamma^{k+1}}{k+1}$ and using $\frac{(\gamma^{k+1})^2}{(k+1)^2}-\frac{\gamma^{k+1}}{k+1}=\frac{(\gamma^k)^2}{k^2}$, we have
		\begin{equation}
		\label{ieq.a6}
		\begin{aligned}
		&\frac{(\gamma^k)^2}{k^2} h^k-\frac{(\gamma^{k+1})^2}{(k+1)^2}h^{k+1}\\
		\ge&
		\sum_{i\in\mathcal{N}_P}\sum_{j\in\mathcal{N}_i}\frac{1}{2\eta_{ij}}\left(\frac{\gamma^{k+1}}{k+1}\left(\lambda^{k+1}_{ij}-\hat{\lambda}^{k+1}_{ij}\right)\right)^2  
		 \\
		&+\sum_{i\in\mathcal{N}_P}\sum_{j\in\mathcal{N}_i}\frac{1}{\eta_{ij}}\frac{\gamma^{k+1}}{k+1}\left( \lambda^{k+1}_{ij}-\hat{\lambda}_{ij}^{k+1}\right)\bigg(\frac{\gamma^{k+1}}{k+1} \hat{\lambda}_{ij}^{k+1}\\
		&-\left(\frac{\gamma^{k+1}}{k+1}-1\right)\lambda_{ij}^{k}-\lambda_{ij}^{*}\bigg).
		\end{aligned}
		\end{equation}
By using $(b-a)^2+2(b-a,a-c) = (b-c)^2 - (a-c)^2$
and combing the inequality (\ref{ieq.a6}) with $a:=\frac{\gamma^{k+1}}{k+1}\hat{\lambda}_{ij}^{k+1},b:=\frac{\gamma^{k+1}}{k+1}\lambda_{ij}^{k+1},c:=(\frac{\gamma^{k+1}}{k+1}-1)\lambda_{ij}^{k} + \lambda_{ij}^{*}$, it has
		\begin{align*}
		&\frac{(\gamma^k)^2}{k^2} h^k-\frac{(\gamma^{k+1})^2}{(k+1)^2}h^{k+1}\\
		&\ge
		\sum_{i\in\mathcal{N}_P}\sum_{j\in\mathcal{N}_i}\frac{1}{2\eta_{ij}}
		\left(\frac{\gamma^{k+1}}{k+1} \lambda_{ij}^{k+1}-\left(\frac{\gamma^{k+1}}{k+1}-1\right)\lambda_{ij}^{k}-\lambda_{ij}^{*}\right) ^2
		\\
		&- 
		\sum_{i\in\mathcal{N}_P}\sum_{j\in\mathcal{N}_i}\frac{1}{2\eta_{ij}}
		\left(\frac{\gamma^{k+1}}{k+1} \hat{\lambda}_{ij}^{k+1}-\left(\frac{\gamma^{k+1}}{k+1}-1\right)\lambda_{ij}^{k}-\lambda_{ij}^{*}\right) ^2.
		\end{align*}
Finally, according to the definition of $\hat{\lambda}^{k+1}$ in Algorithm 1, the result is obtained.
\end{IEEEproof}

\begin{lemma} \label{lemma4}
 If $\{s^k,m^k\}$ are two positive sequences of reals and satisfy
\begin{equation*}
s^k-s^{k+1}\ge m^{k+1}-m^k,\quad\forall k\ge1,\text{with}\ s^1+m^1\le w, w>0,
\end{equation*}
Then for any $k\ge 1$, $s^k\le w$.
\end{lemma}

\begin{lemma} \label{lemma5}
Let the positive sequence $\gamma^k$ be denoted in Algorithm 1 with $\gamma^1=1$.  Then, for all $k\ge1$,
$\gamma^k \ge \frac{k^2}{2}$.
\end{lemma}

In the following, by combining Lemma 3, 4 and 5, the IEEEproof of Theorem 1 is given.

\begin{IEEEproof}
According to Lemma 3 and 4, let $s^k=\frac{(\gamma^k)^2}{k^2} h^k$ and $m^k=\sum_{i\in\mathcal{N}_P}\sum_{j\in\mathcal{N}_i}\frac{1}{2\eta_{ij}}\left(u_{ij}^{k}\right)^2$, then $s^1=(\gamma^1)^2 \cdot h^1 = q(\lambda^1) - q(\lambda^*)$ and $ m^1=\sum_{i\in\mathcal{N}_P}\sum_{j\in\mathcal{N}_i}\frac{1}{2\eta_{ij}}\left(\lambda_{ij}^1-\lambda_{ij}^*\right)^2$.
By using (\ref{ieq.a3}) with the point ($\lambda=\lambda^*,\hat{\lambda}=\hat{\lambda}^1$), it yields
\begin{align*}
s^1+m^1  \le  \sum_{i\in\mathcal{N}_P}\sum_{j\in\mathcal{N}_i}\frac{1}{2\eta_{ij}}\left( \lambda_{ij}^0-\lambda_{ij}^*\right)^2,
\end{align*}
which is due to $\hat{\lambda}_{ij}^1=\lambda_{ij}^0$. By using Lemma 4, it obtains
\begin{align*}
\frac{(\gamma^k)^2}{k^2} h^k \le \sum_{i\in\mathcal{N}_P}\sum_{j\in\mathcal{N}_i}\frac{1}{2\eta_{ij}}\left( \lambda_{ij}^0-\lambda_{ij}^*\right)^2. 
\end{align*}
With $h^k = q(\lambda^k)-q(\lambda^*)$ and Lemma 5, (\ref{pro1}) in Theorem 1 is obtained
\end{IEEEproof}
\section{IEEEproof of Corollary 1}\label{appendix_IEEEproof_corollary1}
\begin{IEEEproof}
By using (\ref{ieq.a3}) with points $(\lambda=\lambda^*, \hat{\lambda}=\lambda^{n})$ and $(\lambda=\lambda^n, \hat{\lambda}=\lambda^{n})$, it obtains
		\begin{equation}
		\begin{aligned}
		\label{ieq.a8}
		q(\lambda^*) - q(\lambda^{n+1}) \ge& \sum_{i\in\mathcal{N}_P}\sum_{j\in\mathcal{N}_i} \frac{1}{2\eta_{ij}}\left(\lambda_{ij}^*-\lambda_{ij}^{n+1}\right)^2\\
		&-
		\sum_{i\in\mathcal{N}_P}\sum_{j\in\mathcal{N}_i}  \frac{1}{2\eta_{ij}}( \lambda_{ij}^{*} - \lambda_{ij}^{n})^2,\\
		\end{aligned}
		\end{equation}
		\begin{equation}
		\label{ieq.a9}
		q(\lambda^n) - q(\lambda^{n+1}) \ge \sum_{i\in\mathcal{N}_P}\sum_{j\in\mathcal{N}_i} \frac{1}{2\eta_{ij}}\left(\lambda_{ij}^{n+1}-\lambda_{ij}^{n}\right)^2.
		\end{equation}
Sum (\ref{ieq.a8}) over $n=0, ..., k-1$. Multiply (\ref{ieq.a9}) by $n$ and sum it over $n=0, ..., k-1$. Then, adding the two results obtained, we can get (\ref{corollary}) in Corollary 1.
\end {IEEEproof}


\bibliographystyle{IEEEtran}
\bibliography{IEEEabrv,mybibfile}

\begin{thebibliography}{10}
\providecommand{\url}[1]{#1}
\csname url@samestyle\endcsname
\providecommand{\newblock}{\relax}
\providecommand{\bibinfo}[2]{#2}
\providecommand{\BIBentrySTDinterwordspacing}{\spaceskip=0pt\relax}
\providecommand{\BIBentryALTinterwordstretchfactor}{4}
\providecommand{\BIBentryALTinterwordspacing}{\spaceskip=\fontdimen2\font plus
\BIBentryALTinterwordstretchfactor\fontdimen3\font minus
  \fontdimen4\font\relax}
\providecommand{\BIBforeignlanguage}[2]{{%
\expandafter\ifx\csname l@#1\endcsname\relax
\typeout{** WARNING: IEEEtran.bst: No hyphenation pattern has been}%
\typeout{** loaded for the language `#1'. Using the pattern for}%
\typeout{** the default language instead.}%
\else
\language=\csname l@#1\endcsname
\fi
#2}}
\providecommand{\BIBdecl}{\relax}
\BIBdecl

\bibitem{2}
Y.~Liu, L.~Wu, and J.~Li, ``Peer-to-peer (p2p) electricity trading in
  distribution systems of the future,'' \emph{Electr. J.}, vol.~32, no.~4, pp.
  2--6, 2019.

\bibitem{4}
Y.~Parag and B.~K. Sovacool, ``Electricity market design for the prosumer
  era,'' \emph{Nat. Energy}, vol.~1, no.~4, pp. 1--6, 2016.

\bibitem{48}
C.~Liu and Z.~Li, ``Comparison of centralized and peer-to-peer decentralized
  market designs for community markets,'' \emph{{IEEE} Trans. Ind. Appl.},
  vol.~58, no.~1, pp. 67--77, 2022.

\bibitem{10}
W.~Tushar, T.~K. Saha, C.~Yuen, D.~Smith, and H.~V. Poor, ``Peer-to-peer
  trading in electricity networks: An overview,'' \emph{IEEE Trans. Smart
  Grid}, vol.~11, no.~4, pp. 3185--3200, 2020.

\bibitem{51}
A.~A. Raja and S.~Grammatico, ``Bilateral peer-to-peer energy trading via
  coalitional games,'' \emph{{IEEE} Trans. Ind. Informat.}, vol.~19, no.~5, pp.
  6814--6824, 2023.

\bibitem{47}
K.~Chen, J.~Lin, and Y.~Song, ``Trading strategy optimization for a prosumer in
  continuous double auction-based peer-to-peer market: A prediction-integration
  model,'' \emph{Appl. Energy}, vol. 242, pp. 1121--1133, 2019.

\bibitem{22}
G.~Hug, S.~Kar, and C.~Wu, ``Consensus+ innovations approach for distributed
  multiagent coordination in a microgrid,'' \emph{IEEE Trans. Smart Grid},
  vol.~6, no.~4, pp. 1893--1903, 2015.

\bibitem{50}
A.~Paudel, M.~Khorasany, and H.~B. Gooi, ``Decentralized local energy trading
  in microgrids with voltage management,'' \emph{{IEEE} Trans. Ind. Informat.},
  vol.~17, no.~2, pp. 1111--1121, 2021.

\bibitem{49}
M.~H. Ullah and J.-D. Park, ``A two-tier distributed market clearing scheme for
  peer-to-peer energy sharing in smart grid,'' \emph{{IEEE} Trans. Ind.
  Informat.}, vol.~18, no.~1, pp. 66--76, 2022.

\bibitem{A-1}
E.~Sorin, L.~Bobo, and P.~Pinson, ``Consensus-based approach to peer-to-peer
  electricity markets with product differentiation,'' \emph{{IEEE} Trans. Power
  Syst.}, vol.~34, no.~2, pp. 994--1004, 2019.

\bibitem{42}
D.~H. Nguyen, ``Optimal solution analysis and decentralized mechanisms for
  peer-to-peer energy markets,'' \emph{{IEEE} Trans. Power Syst.}, vol.~36,
  no.~2, pp. 1470--1481, 2021.

\bibitem{31}
T.~Morstyn and M.~D. McCulloch, ``Multiclass energy management for peer-to-peer
  energy trading driven by prosumer preferences,'' \emph{{IEEE} Trans. Power
  Syst.}, vol.~34, no.~5, pp. 4005--4014, 2019.

\bibitem{33}
T.~Baroche, P.~Pinson, R.~L.~G. Latimier, and H.~B. Ahmed, ``Exogenous cost
  allocation in peer-to-peer electricity markets,'' \emph{{IEEE} Trans. Power
  Syst.}, vol.~34, no.~4, pp. 2553--2564, 2019.

\bibitem{38}
A.~Paudel, L.~Sampath, J.~Yang, and H.~B. Gooi, ``Peer-to-peer energy trading
  in smart grid considering power losses and network fees,'' \emph{IEEE Trans.
  Smart Grid}, vol.~11, no.~6, pp. 4727--4737, 2020.

\bibitem{A-2}
M.~Khorasany, Y.~Mishra, and G.~Ledwich, ``A decentralized bilateral energy
  trading system for peer-to-peer electricity markets,'' \emph{{IEEE} Trans.
  Ind. Electron.}, vol.~67, no.~6, pp. 4646--4657, 2020.

\bibitem{28}
K.~Umer, Q.~Huang, M.~Khorasany, M.~Afzal, and W.~Amin, ``A novel communication
  efficient peer-to-peer energy trading scheme for enhanced privacy in
  microgrids,'' \emph{Appl. Energy}, vol. 296, 2021, {A}rt. no. 117075.

\bibitem{13}
M.~Khorasany, Y.~Mishra, B.~Babaki, and G.~Ledwich, ``Enhancing scalability of
  peer-to-peer energy markets using adaptive segmentation method,'' \emph{J.
  Mod. Power Syst. Clean Energy}, vol.~7, no.~4, pp. 791--801, 2019.

\bibitem{34}
A.~Mas-Colell, M.~D. Whinston, and J.~R. Green, \emph{Microeconomic Theory, 1st
  ed}.\hskip 1em plus 0.5em minus 0.4em\relax Oxford University Press, 1995.

\bibitem{15}
P.~Samadi, H.~Mohsenian-Rad, R.~Schober, and V.~W. Wong, ``Advanced demand side
  management for the future smart grid using mechanism design,'' \emph{IEEE
  Trans. Smart Grid}, vol.~3, no.~3, pp. 1170--1180, 2012.

\bibitem{43}
M.~H. Ullah and J.-D. Park, ``Peer-to-peer energy trading in transactive
  markets considering physical network constraints,'' \emph{IEEE Trans. Smart
  Grid}, vol.~12, no.~4, pp. 3390--3403, 2021.

\bibitem{41}
J.~Guerrero, A.~C. Chapman, and G.~Verbič, ``Decentralized p2p energy trading
  under network constraints in a low-voltage network,'' \emph{IEEE Trans. Smart
  Grid}, vol.~10, no.~5, pp. 5163--5173, 2019.

\bibitem{26}
D.~Bertsekas, \emph{Nonlinear Programming}.\hskip 1em plus 0.5em minus
  0.4em\relax Athena Scientific, 1995.

\bibitem{25}
S.~Boyd, N.~Parikh, and E.~Chu, \emph{Distributed optimization and statistical
  learning via the alternating direction method of multipliers}.\hskip 1em plus
  0.5em minus 0.4em\relax Now Publishers Inc, 2011.

\bibitem{39}
N.~G. Mankiw, \emph{Principles of microeconomics}.\hskip 1em plus 0.5em minus
  0.4em\relax Cengage Learning, 2011.

\bibitem{32}
A.~Nedi{\'c}, \emph{LectureNotes Optimization I}.\hskip 1em plus 0.5em minus
  0.4em\relax Hamilton Institute, 2008.

\end{thebibliography}

\end{document}